\documentclass[11pt]{article}

\def\mytitle#1{\setcounter{equation}{0}
\setcounter{footnote}{0}
\begin{flushleft}\Large\textbf{#1}\end{flushleft}
\vspace{0.25cm}}
\def\myname#1{\leftline{{\large #1}}\vspace{-0.13cm}}
\def\myplace#1#2{\small\begin{flushleft}\textit{#1}\\
\texttt{#2}\end{flushleft}}

\def\myclassification#1{\small\noindent
Pacs no :
       #1\vspace{0.5cm}}
\usepackage{graphicx}
\begin{document}

\mytitle{Interacting Holographic Dark Energy at the Ricci scale and Dynamical system}

\vskip0.2cm \myname{Nairwita
Mazumder\footnote{nairwita15@gmail.com}}
\vskip0.2cm\myname{Ritabrata
Biswas\footnote{biswas.ritabrata@gmail.com}} \vskip0.2cm
\myname{Subenoy
Chakraborty\footnote{schakraborty@math.jdvu.ac.in}}

\myplace{Department of Mathematics, Jadavpur University,
Kolkata-700 032, India.} { }

\begin{abstract}
In this work, we consider homogeneous and isotropic FRW model of
the universe, filled with interacting dark matter and dark energy.
The dark matter is chosen as usual in the form of dust while dark
energy is holographic in nature with IR cut off at the Ricci's
length and it is in the form of a perfect fluid with variable
equation of state. We have chosen the interaction term of the
following two types: (i) a linear combination of the matter
density of the two fluids, (ii) a product of the two matter
densities. For both the choices the evolution equations are
transformed to an autonomous system and the corresponding critical
points are analyzed. Finally, for the first choice of the
interaction term the evolution of the ratio of the energy
densities has been studied from the point of view of the present
coincidence problem.\\

\end{abstract}
\myclassification{04.20.-q, 04.40.-b, 95.35.+d, 98.80.Cq}
\section{Introduction}
In the last decade there are wide variety of modern cosmological
observations namely precision measurements of anisotropies in the
cosmological microwave background radiation \cite{Spergel1},
baryon acoustic oscillation \cite{Pereival1,Einstein1} and type Ia
supernova \cite{Riess1,Perlmutter1,Amanullah1}. These observations
indicate that at present our universe is composed of nearly $25\%$
cold dark matter (DM), $70\%$ nonbaryonic unknown matter known as
dark energy
(DE)\cite{Riess1,Perlmutter1,Perlmutter2,Tonry1,Riess2}and $5\%$
of radiation and baryonic matter which ia well understood by the
standard models of particles. The natural and leading choice of
the unknown DE is the cosmological constant($\Lambda CDM$ model)
which represents a vacuum energy density having constant equation
of state $\omega=-1$. However, its observed value is far below
than the esteemation from quantum field theory(known as
cosmological constant problem). Also there is no expectation why
the constant vacuum energy and matter energy densities are
precisely of the same order today(coincidence problem). Due to
these observational \cite{Peebles1} and theoretical
\cite{Weinberg1,Copeland1} probes for the cosmological constant
there are alternative models for DE (varies from time) in the
literature. Scalar field models \cite{Ratra1,Caldwell1,Copeland2}
(known as quintesence) have attracted special attention compared
to the other alternatives \cite{Copeland1}.

The fact that at present DE and DM are dominant sources of the
content of the universe, there has been a lot of interest in
studying coupling in the dark sector components
\cite{Pavon1,Mangano1,Micheletti1,Sandro1,Amendola1,Pavon2,
Boehmer1,Olivares1,Chen1,Hsu1,Hooft1,Cohen1,Gao1,Xu1,
Suwa1,Lepe1,Brustein}. It is partly motivated by the fact that one
can only extract information of these components through
gravitational interaction. Also consideration of interaction is
natural in the framework of field theory
\cite{Micheletti1,Sandro1}. Recently, it has been shown that an
appropriate choice of the interaction between DE and DM can
alleviate the coincidence problem
\cite{Amendola1,Pavon2,Boehmer1,Olivares1,Chen1}.

To have some inside about the unknown and mysterious nature of DE,
many people have suggested that DE should be compatible with
Holographic principle ,namely "the number of relevant degrees of
freedom of a system dominated by gravity must vary along with the
area of the surface bounding the system"\cite{Hooft1}. Such a DE
model is known as Holographic DE(HDE) model. Further the energy
density of any given region should be bound by that ascribed to a
Schwarzschild black hole(BH) that fills the same volume
\cite{Cohen1}. Mathematically, we write $\rho_{D}\leq
M_{p}^{2}L^{-2}$, where $\rho_{D}$ is the DE density, $L$ is the
size of the region(or infrared cut off) and $M_{p}=\left(8\pi
G\right)^{-\frac{1}{2}}$ is the reduced Planck mass. Usually, the
DE density is written as
\begin{equation}\label{1}
\rho_{D}=\frac{3M_{p}^{2}c^{2}}{L^{2}}
\end{equation}
Here the dimensionless parameter '$c^2$' takes care of the
uncertainties of the theory and for mathematical convenience the
factor 3 has been introduced. Due to lack of clear idea there are
many choices for the infrared cut off of which the most relevant
one are the Hubble radius,i.e., $L=H^{-1}$\cite{Pavon2,Hsu1} and
the Ricci's length, i.e.,
$L=\left(\dot{H}+2H^{2}\right)^{\frac{-1}2}$
\cite{Gao1,Xu1,Suwa1,Lepe1}. The argument behind the choice of
Ricci's length as the IR cut-off is that it corresponds to the
size of the maximal perturbation, leading to the formation of a
black hole\cite{Brustein}. Another commonly used IR cut-off length
is the radius of future event horizon, but it suffers from a
severe circularity problem.

In the present paper we consider a cosmological model of
Holographic DE (HDE) in the form of a perfect fluid interacting
with DM in the form of dust. The choice of the IR cut-off is
chosen as the Ricci's length,i.e.,\cite{Duran1}
\begin{equation}\label{2}
L=\left(\dot{H}+2H^{2}\right)^{-\frac{1}{2}}
\end{equation}
The evolution equations of the model are formulated into an
autonomous system and critical points are analysed. Explicit
solutions are obtained and are analysed asymptotically.

\section{Basic equations for interacting HDE at the Ricci scale}\label{chapter2}
In the present work, the homogeneous and isotropic FRW universe is
assumed to fill up with interacting two fluid system-one component
is in the form of dust(having energy density $\rho_{m}$) known as
dark matter(DM) while perfect fluid having barotropic equation of
state $p_{D}=\omega_{D}\rho_{D}$, $\omega_{D}$, a variable is the
DE component.

Assuming spatialy flat model, the friedmann equations are (choosing $8\pi G =1=c$)
\begin{equation}\label{3}
3H^{2}=\rho_{m}+\rho_{D}
\end{equation}
and
\begin{equation}\label{4}
2\dot{H}=-\rho_{m}-\left(1+\omega_{D}\right)\rho_{D}
\end{equation}
and we have the conservation equations
\begin{equation}\label{5}
\dot{\rho}_{m}+3H\rho_{m}=Q
\end{equation}
\begin{equation}\label{6}
\dot{\rho}_{D}+3H\left(1+\omega_{D}\right)\rho_{D}=-Q
\end{equation}
Here the interaction term $Q>0$ indicates transfer of energy from
DE component to DM sector while opposite is the situation for
$Q<0$. As $Q<0$ would worsen the coincidence problem so we choose
$Q>0$ throughout the work. Also validity of the second law of
thermodynamics and Le chatelier's principle \cite{Pavon1,Lip1}
support this choice of positive Q. It should be noted that we have
not included baryonic matter in the interaction due to the
constraints imposed by local gravity measurements
\cite{Ratra1,Lip1,Hagiwara1}. In the next two sections we shall
deal with two different choices of interaction term separately,
namely, $(i)$ $Q=3b^{2}H \rho$ ($\rho=\rho_{m}+\rho_{D}$, the
total energy density) and $(ii)$ $Q=\gamma
\rho_{m}\rho_{D}~(\gamma>0)$. The first choice is the special case
of the usual one used in the literature as a linear combination of
the energy densities. The second choice is physically more viable
in the sense that interaction rate vanishes if one of the
densities is zero and increases with each of the densities. Also
this choice of interaction for HDE models gives the best fit to
observations \cite{Pavon1,Lip1}. Also it should be noted that the
constant $\gamma$ has the dimension of $\left[L^{3}MT\right]$.\\

Using the field equations (\ref{3}) and (\ref{4}) we have the form (\ref{1}) the expression for the energy density of HDE as
\begin{equation}\label{7}
\rho_{D}=\frac{c^{2}}{2}\left\{\rho_{m}+\left(1-3\omega_{D}\right)\rho_{D}\right\}
\end{equation}
Hence the equation of state parametercan be expressed in terms of the density parameter as
\begin{equation}\label{8}
\omega_{D}=-\frac{2}{3c^{2}}+\frac{1}{3\Omega_{D}}
\end{equation}
Also the deceleration parameter takes the simple form
\begin{equation}\label{9}
q=-\frac{\ddot{a}}{a^{2}H^{2}}=1-\frac{\Omega_{D}}{c^{2}}=1-\frac{1}{c^{2}\left(1+u\right)}
\end{equation}
which shows a smooth transition from deceleration to acceleration
as universe evolves from the early matter dominated era to the
late time DE dominated and here
$u=\frac{\rho_m}{\rho_D}=\frac{1}{\Omega_{D}}-1$.
\section{Explicit calculations for choice of interaction term}\label{chapter3}

{\bf Case (I) : $Q=3b^{2}H\rho$}\\

The energy conservation equations can be written explicitly as
\begin{equation}\label{10}
\dot{\rho}_{m}=\sqrt{3\left(\rho_{m}+\rho_{D}\right)}\left[b^{2}\rho_{D}-\left(1-b^{2}\right)\rho_{m}\right]
\end{equation}
\begin{equation}\label{11}
\dot{\rho}_{D}=-3H \left[b^{2}\left(\rho_{m}+\rho_{D}\right)+
\rho_{D}\left(1+\omega_{D}\right)\right]=-\sqrt{3\left
(\rho_{m}+\rho_{D}\right)}\left[b^{2}\left(\rho_{m}+\rho_{D}\right)+\left(1+\frac{1}{3\Omega_{D}}-\frac{2}{3c^{2}}\right)\rho_{D}\right]
\end{equation}
As a consequence the evolution of the density parameter
$\Omega_{D}$ and the ratio of the energy densities $u$ has the
form
\begin{equation}\label{12}
\dot{\Omega}_{D}=-H\left[-\frac{2\Omega_{D}\left(1-\Omega_{D}\right)}{c^{2}}+\left(1-\Omega_{D}\right)+3b^{2}\right]
\end{equation}
\begin{equation}\label{13}
\dot{u}=H \left[-\frac{2u}{c^{2}}+u\left(1+u\right)+3b^{2}\left(1+u\right)^{2}\right]
\end{equation}

\begin{figure}
\includegraphics[height=2.6in, width=2.6in]{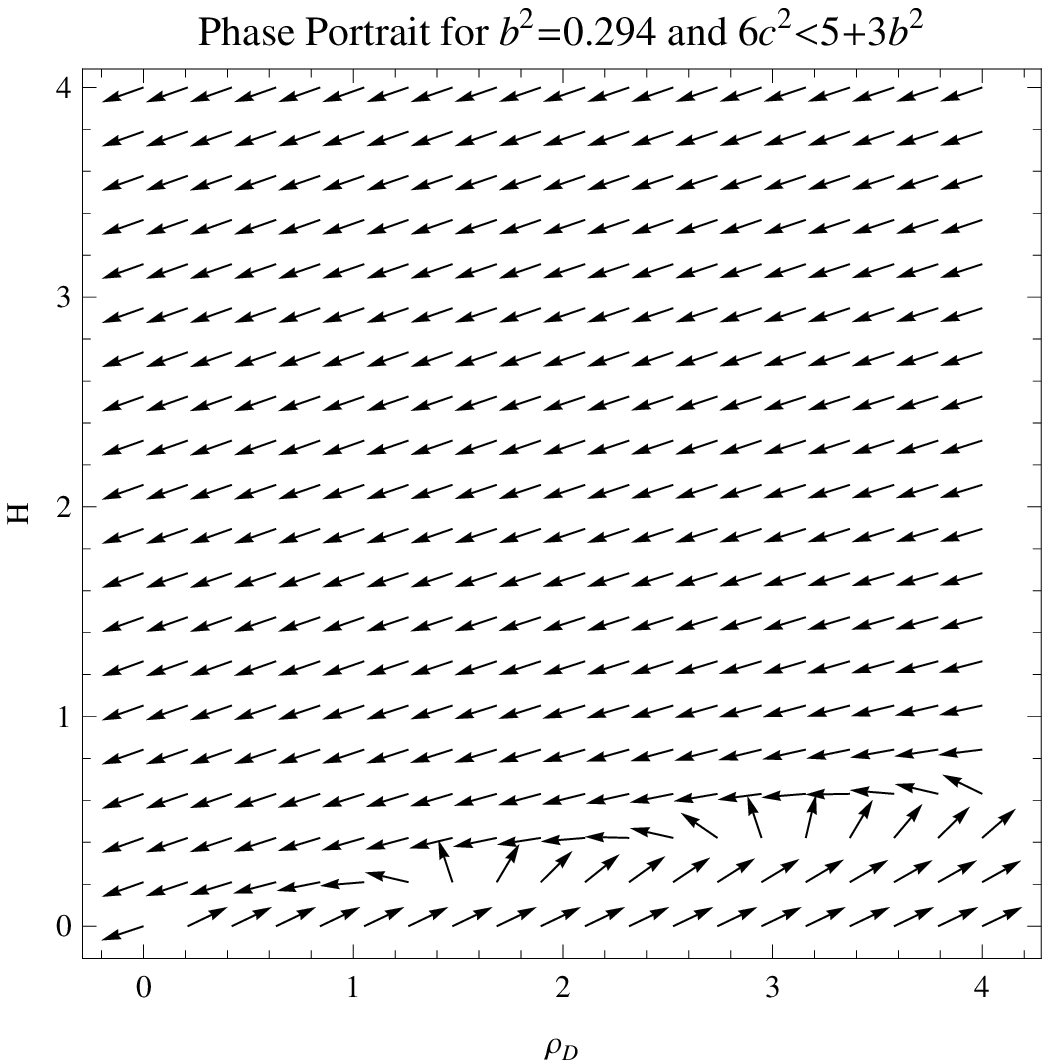}~~
\includegraphics[height=2.6in, width=2.6in]{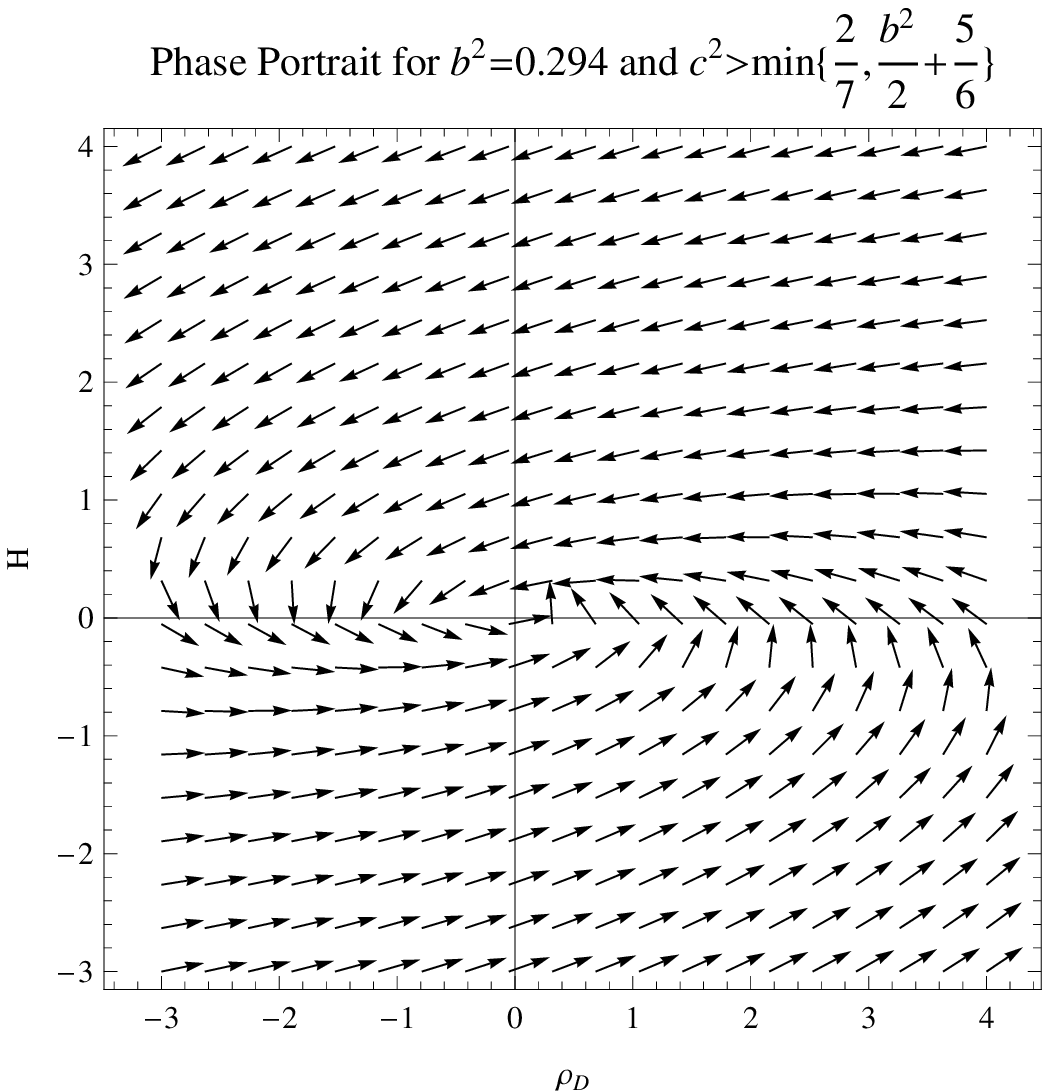}\\
~~~~~~~~~~~~~~~~~~~~~~~~~~~~~~~~~~~~~~~~~~~~~~~~~~~~~~~~~~~~~~~~~Fig.1(a)
.~~~~~~~~~~~~~~~~~~~~~~~~~~~~~~~~~~~~~~~~~~~~~~~~~~~~~~~~~~~~~~~~Fig.1(b)
\\\\

Fig. 1(a)-1(b) represent the variation of $\rho_D - H$. Though in
Fig 1(b) the whole region is not a physically valid region but for
better understanding about the system and the nature of the
critical point we have drawn graph for the whole region.

\hspace{1cm} \vspace{2cm}

\end{figure}

From equation (\ref{12}) we see that the DE density gradually
decreases with the evolution of the universe until it is in the
phantom era $(1+\omega_{D}<0)$. Thus if we assume the DE density
to be sufficiently large at the early epochs of the evolution then
from equation (\ref{11}) $\dot{\rho_{m}}=0$ increases till some
intermediate stage and then gradually decreases with
$\dot{\rho}_{m}=0$ along the line
$\frac{\rho_{m}}{\rho_{D}}=\frac{b^{2}}{1-b^{2}}$ in the
$\left(\rho_{m},~\rho_{D}\right)$ plane. Thus we have $u\sim O(1)$
at some intermediate stage of evolution in the neighbourhood of
the above line and may be a possible resolution of coincidence
problem \cite{Duran1,Lip1}. Further, the present model of the
universe shows a DE dominance at the early epochs and subsequently
universe evolves with DM as the dominant component and then again
it has DE dominated phase at late time as predicted by
observation. Hence
the above model is suitable for the present universe.\\

Now to formulate an autonomous system we rewrite equation
(\ref{12}) using field equation (\ref{3}) as
\begin{equation}\label{14}
\rho_{D}'=-\left[3H^{2}\left(1+3b^{2}\right)+\rho_{D}\left(3-\frac{2}{c^{2}}\right)\right]
\end{equation}
and the second field equation, i.e., equation (\ref{4}) can be written as
\begin{equation}\label{15}
\left({H}^{2}\right)'=-2\left[2H^{2}-\frac{\rho_{D}}{3c^{2}}\right]
\end{equation}

where $'\equiv \frac{\partial}{\partial x},~x=ln a$ Thus equations
(\ref{14}) and (\ref{15}) form a linear homogeneous autonomous
system in the phase plane $\left(\rho_{D},~H^{2}\right)$, having
critical point at the origin. For the Jacobi matrix $A$,
$Tr(A)=-7+\frac{2}{c^{2}}$ and
$det(A)=\frac{2}{c^{2}}\left(6c^{2}-5-3b^{2}\right)$.\\

The nature of the critical points is characterized \cite{Perko1}
in the \t{Table1} and the geometrical features are presented in
fig 1(a)
and 1(b).\\

\begin{table}[t]\label{Table1}
\begin{center}
\caption{{\bf Nature~ of~ the~ Critical~Points~for~Case-I}}
\centering
\begin{tabular}{|l|l|l|}
\hline
Condition &Nature of the eigen values & Type of critical point \\

\hline \hline
$(i)$ $6c^{2}<5+3b^{2}$ & $(i)$ real roots of opposite sign & $(i)$ Saddle\\
$(ii)$ $c^{2}>min \left (\frac{2}{7},~\frac{b^2}{2}+\frac{5}{6}\right)$ & $(ii)$ both negative real roots & $(ii)$ Stable nodes.\\

\hline
\end{tabular}
\end{center}
\end{table}

From equation (\ref{13}) if $u_{f}$ be a fixed point, i.e., $\left.\dot{u}\right|_{u=u_{f}}=0$ then the parameter $b^{2}$ has the expression
\begin{equation}\label{16}
b^{2}=\frac{u_{f}}{3}\frac{\left[\frac{2}{c^{2}}-1-u_{f}\right]}{\left(1+u_{f}\right)^{2}}
\end{equation}

Using this value of $b^{2}$ the other fixed point of equation
(\ref{13}) can be expressed in terms of $u_{f}$ as

\begin{equation}\label{17}
u_{p}=\frac{\frac{2}{c^{2}}-\left(1+u_{f}\right)}{\frac{2}{c^2}{u_{f}}+\left(1+u_{f}\right)}
\end{equation}
Further from equation (\ref{13}), differentiating once we obtain
\begin{equation}\label{18}
\frac{du'}{du}=-\frac{2}{c^{2}}+1+2u+6b^{2}\left(1+u\right)
\end{equation}

Now if we assume the fixed points $u_{p}$ and $u_{f}$ to be at the
far past and at the far future respectively, then
$u_{f}<u_{0}\simeq 0.45$ and $c^{2}<\Omega_{D_{0}}\simeq0.75$
\cite{Duran1}, where the suffix $'0'$ stands for the present value
and the second inequality is obtained from equation  (\ref{9})
with the fact that we are at present in an accelerating phase.
Then from (\ref{18})

$$\left.\frac{du'}{du}\right|_{u=u_{p}} = -\frac{\left(2+c^{2}\right)}{c^{2}}+\frac{4}{c^{2}\left(1+u_{f}\right)}>0$$

and
$$\left.\frac{du'}{du}\right|_{u=u_{f}}=1+\frac{2\left(u_{f}-1\right)}{c^{2}\left(u_{f}+1\right)}<0$$
So the fixed point at far past is an unstable one while the fixed
point at far future is a stable one.\\
Moreover, integrating equation (\ref{13}) we obtain
\begin{equation}\label{19}
u=\frac{\left(u_{p}+u_{f}\right)}{2}+\frac{\left(u_{f}-u_{p}\right)}{2}\left[\frac{1+k(\tilde{a})^{\mu}}{k(\tilde{a})^{\mu}-1}\right]
\end{equation}
where
$k=\frac{u_{0}-u_{p}}{u_{0}-u_{f}},~\tilde{a}=\frac{a}{a_{0}}~and
~ \mu=\left(1+3b^{2}\right)\left(u_{p}-u_{f}\right)$.

Further, equation (\ref{13}) can be expressed in terms of $u_{p}$
and $u_{f}$ as
\begin{equation}\label{20}
\dot{u}=H\left(1+3b^{2}\right)\left(u-u_{p}\right)\left(u-u_{f}\right),
\end{equation}

which clearly shows that $\dot{u}<0$ between the two fixed points
$u_{p}$ and $u_{f}$. The continuous decrease of $u$ between the
two fixed points is shown in figure {\bf $2(a)$} where $u\approx
1$ near $\tilde{a}=0.8$. Hence the coincidence problem has some
partial solution for the present model, it can not predict
$u_{0}\sim O(1)$ \cite{Duran1}. The explicit expression for the
density parameter $\Omega_{D}$ is given by
\begin{equation}\label{21}
\Omega_{D}=\frac{k(\tilde{a})^{\mu}-1}{\left(1+u_{f}\right)k(\tilde{a})^{\mu}-\left(1+u_{p}\right)}
\end{equation}
and hence using (\ref{8}) we have

\begin{equation}\label{22}
\omega_{D}=-\frac{2}{3c^{2}}+\frac{1}{3}\frac{\left(1+u_{f}\right)k(\tilde{a})^{\mu}-\left(1+u_{p}\right)}{k(\tilde{a})^{\mu}-1}
\end{equation}

\begin{figure}
\includegraphics[height=2.6in, width=2.6in]{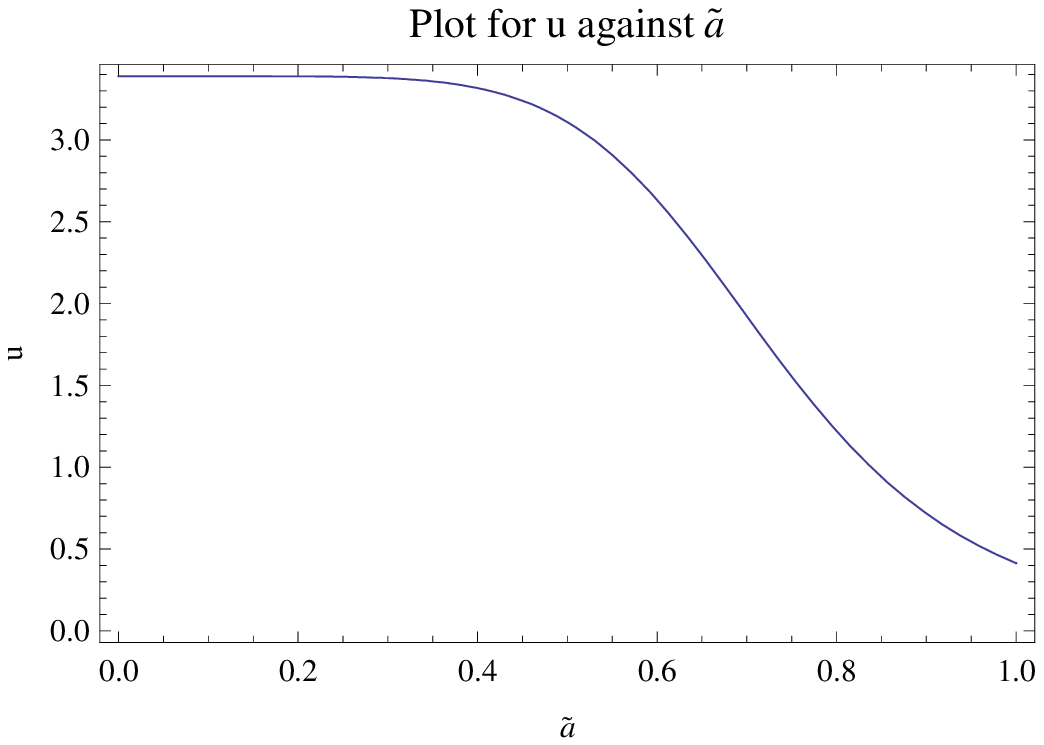}~~
\includegraphics[height=2.6in, width=2.6in]{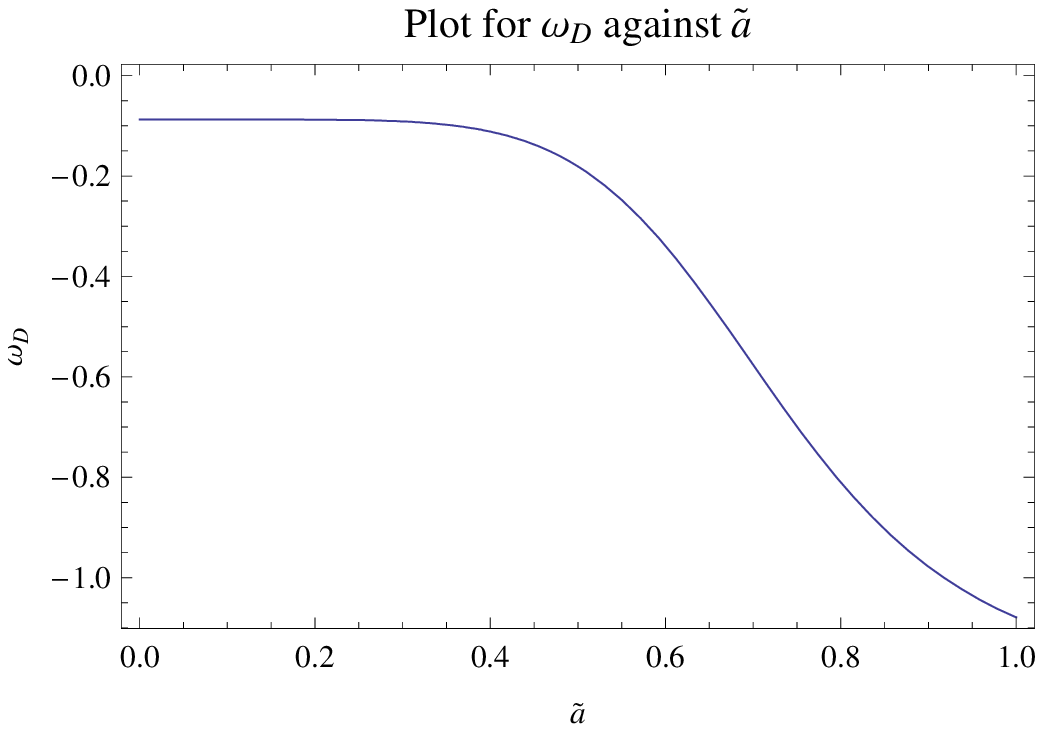}\\
~~~~~~~~~~~~~~~~~~~~~~~~~~~~~~~~~~~~~~~~~~~~~~~~~~~~~~~~~~~~~~~~~Fig.2(a)
.~~~~~~~~~~~~~~~~~~~~~~~~~~~~~~~~~~~~~~~~~~~~~~~~~~~~~~~~~~~~~~~~Fig.2(b)
\\\\

Fig. 2(a)-2(b) represent the variation of $u$ and $\omega_D$
against $\tilde{a}$ respectively corresponding to the value of the
parameters $u_f=0.013,c^2=0.44,u_0 = 0.4144~ and~  b^2 = 0.290$.

\hspace{1cm} \vspace{2cm}

\end{figure}

The variation of $\omega_{D}$ over the scale factor is shown in
figure 2(b) which shows that we are very close to $\Lambda CDM$
era in the present epoch. At the two extreme limits the limiting
values of $\omega_{D}$ are as follows :
$$\omega_{D}\rightarrow \omega_{D0}=-\frac{2}{3c^{2}}+\frac{\left(1+u_{p}\right)}{3}~~~~~~(a\rightarrow 0)$$
$$\omega_{D}\rightarrow \omega_{\infty}=-\frac{2}{3c^{2}}+\frac{\left(1+u_{p}\right)}{3}~~~~~~(a\rightarrow \infty)$$

Note that as $u$ decreases with the evolution so $u_{f}<u_{p}$ and
hence $\omega_{D}$ also decreases as the universe grows up. If we
choose $u_{0}\simeq 1$ then the present value of $\omega_{D}$
(i.e.,$\omega_{D0}$) does not depend on the asymptotic values
$u_{p}$ and $u_{f}$, i.e.,
$$\omega_{D0}=-\frac{2}{3}\left(-1+\frac{1}{c^{2}}\right)$$
which is compatible with recent observation, i.e., $\omega_{D0}$
should be very close to $-1$ (as shown in Fig.2(b)) if
$c^{2}\simeq 0.4$ and is closed to the estimated lower bound of
$c^{2}$. Integrating field equation (\ref{4}) using equation
(\ref{21}) and (\ref{22}) we have
\begin{equation}\label{23}
H=H_{0}(\tilde{a})^{\left\{\frac{1}{c^{2}\left(1+u_{p}\right)}-2\right\}}\times\left[\frac{\left(1+u_{f}\right)k\tilde{a}^{\mu}-\left(1+u_{p}\right)}{\left(1+u_{f}\right)k-\left(1+u_{p}\right)}\right]^{\frac{1}{2}}
\end{equation}
Now combining equation (\ref{21}) and (\ref{23}) the HDE density
has the form
\begin{equation}\label{24}
\rho_{D}=\frac{3H_{0}^{2}}{\left(u_{p}-u{f}\right)\left(u_{0}+1\right)}\left[\left(u_{p}-u_{0}\right)(\tilde{a})^{2\left\{\frac{1}{c^{2}\left(1+u_{f}\right)}-2\right\}}+\left(u_{0}-u_{f}\right)(\tilde{a})^{2\left\{\frac{1}{c^{2}\left(1+u_{p}\right)}-2\right\}}\right]
\end{equation}
The above expression for $\rho_{D}$ contains two terms $-$ the
first one is dominant at later epochs when $a$ is large while the
second term is the dominant one at early phases.\\

{\bf Case (II) : $Q=\gamma \rho_{m} \rho_{D}$}\\

As before the explicit form of the energy conservation equations
are
\begin{equation}\label{25}
\dot{\rho}_{m}=\rho_{m}\left[\gamma \rho_{D}-3H\right]
\end{equation}

and

\begin{equation}\label{26}
\dot{\rho}_{D}=-\rho_{D}\left[\gamma \rho_{m}+3H\left(1+\omega_{D}\right)\right]
\end{equation}

and hence the ratio of the energy densities has the evolution
equation

\begin{equation}\label{27}
\dot{u}=3H u \left[\gamma H+\omega_{D}\right]=3H u \left[\gamma
H+\frac{(1+u)}{3}-\frac{2}{3c^{2}}\right]
\end{equation}
The field equation (\ref{4}) can be written as
\begin{equation}\label{28}
\dot{H}=-\frac{3H^{2}}{2}\left[\frac{4}{3}-\frac{2}{3c^{2}(1+u)}\right]
\end{equation}
Thus equations (\ref{27}) and (\ref{28}) from an autonomous system
in the $(u,~H)$-phase plane. The only critical point which is of
physical interest is $\left(\frac{1}{2c^{2}}-1,~\frac{1}{2\gamma
c^{2}}\right)$ in the $(u,~H)$ phase plane. It is to be noted that
the other critical points correspond to static model of the
universe or a degenerate line \cite{Perko1} representing universe
filled up with DE only. The critical point is characterized in the
\t{Table2} where for the linearized matrix $A$,
$p=tr(A)=\frac{1}{2c^{4}\gamma}\left(\frac{1}{2}-c^{2}\right)$,
$q=det A=\frac{3}{2c^{6}\gamma^{2}}\left(\frac{1}{2}-c^{2}\right)$
and
$\Delta=p^{2}-4q=\frac{25}{4c^{8}\gamma^{2}}\left(\frac{1}{2}-c^{2}\right)\left(\frac{1}{50}-c^{2}\right)$.
The geometric nature of the equilibrium point in each of the above
five cases are presented in figures {\bf 3(a)-(e)}.

\begin{table}[t]\label{Table2}
\begin{center}
\caption{{\bf Nature~ of~ the~ Critical~Points~for~Case-II}}
\centering

\begin{tabular}{|l|l|l|}
\hline
Condition &Nature of the eigen values & Type of critical point \\

\hline \hline
$(i)$ $\frac{1}{50}<c^{2}<\frac{1}{2}$ & $(i)$ pair of complex roots with real part positive. & $(i)$ unstable spiral.\\
$(ii)$ $c^{2}>\frac{1}{2}$ & $(ii)$  real roots of opposite sign.  & $(ii)$ Saddel.\\
$(iii)$ $c^{2}<\frac{1}{50}$ & $(iii)$ real roots of same sign. & $(iii)$ Unstable node.\\
$(iv)$ $c^{2}=\frac{1}{50}$ & $(iii)$ equal non-zero real roots. & $(iii)$ degenerate unstable node.\\
$(v)$ $c^{2}=\frac{1}{2}$ & $(iii)$ zero roots. & $(iii)$ degenerate point.\\
\hline
\end{tabular}
\end{center}
\end{table}

\begin{figure}
\includegraphics[height=2.6in, width=2.6in]{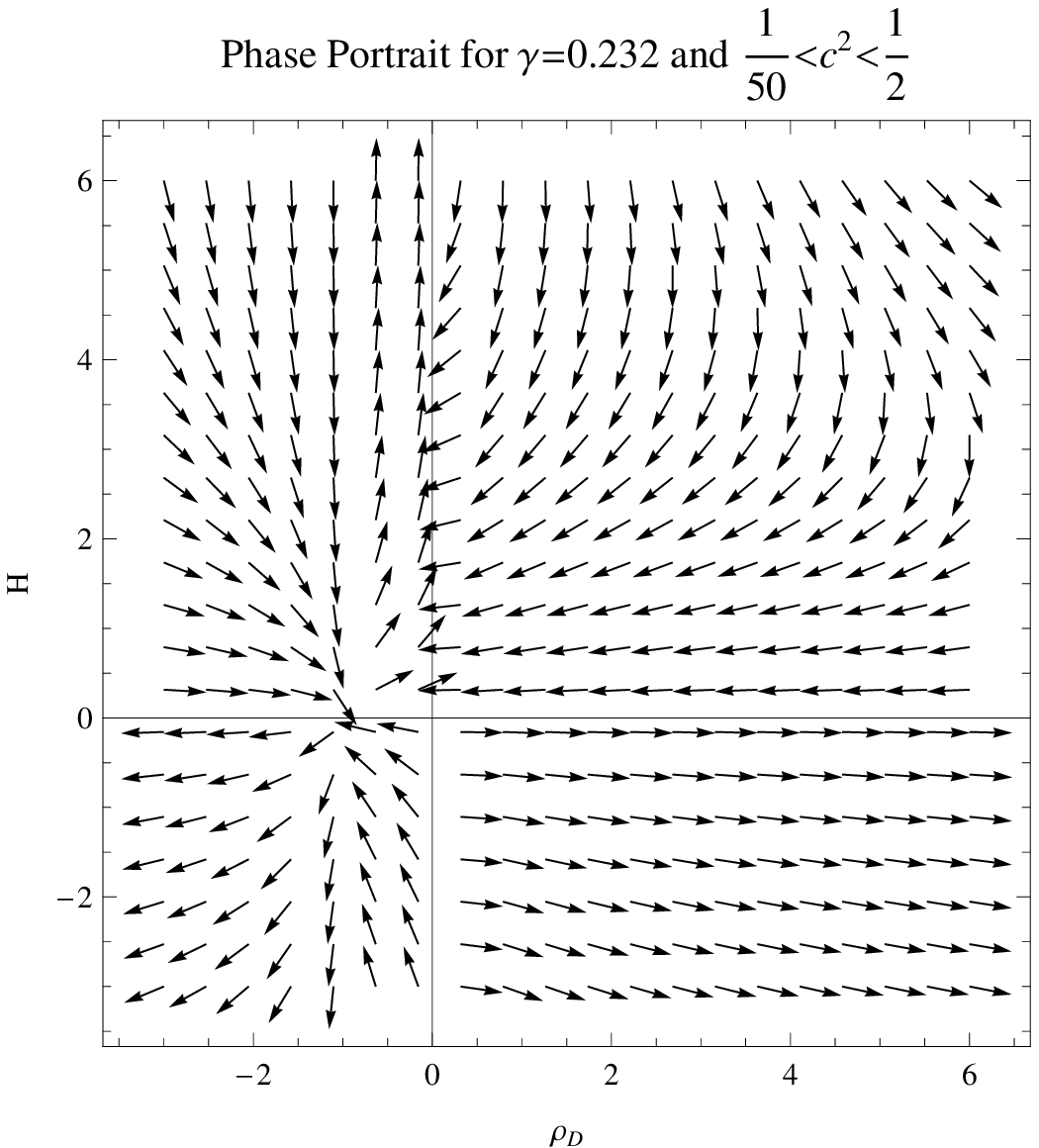}~~
\includegraphics[height=2.6in, width=2.6in]{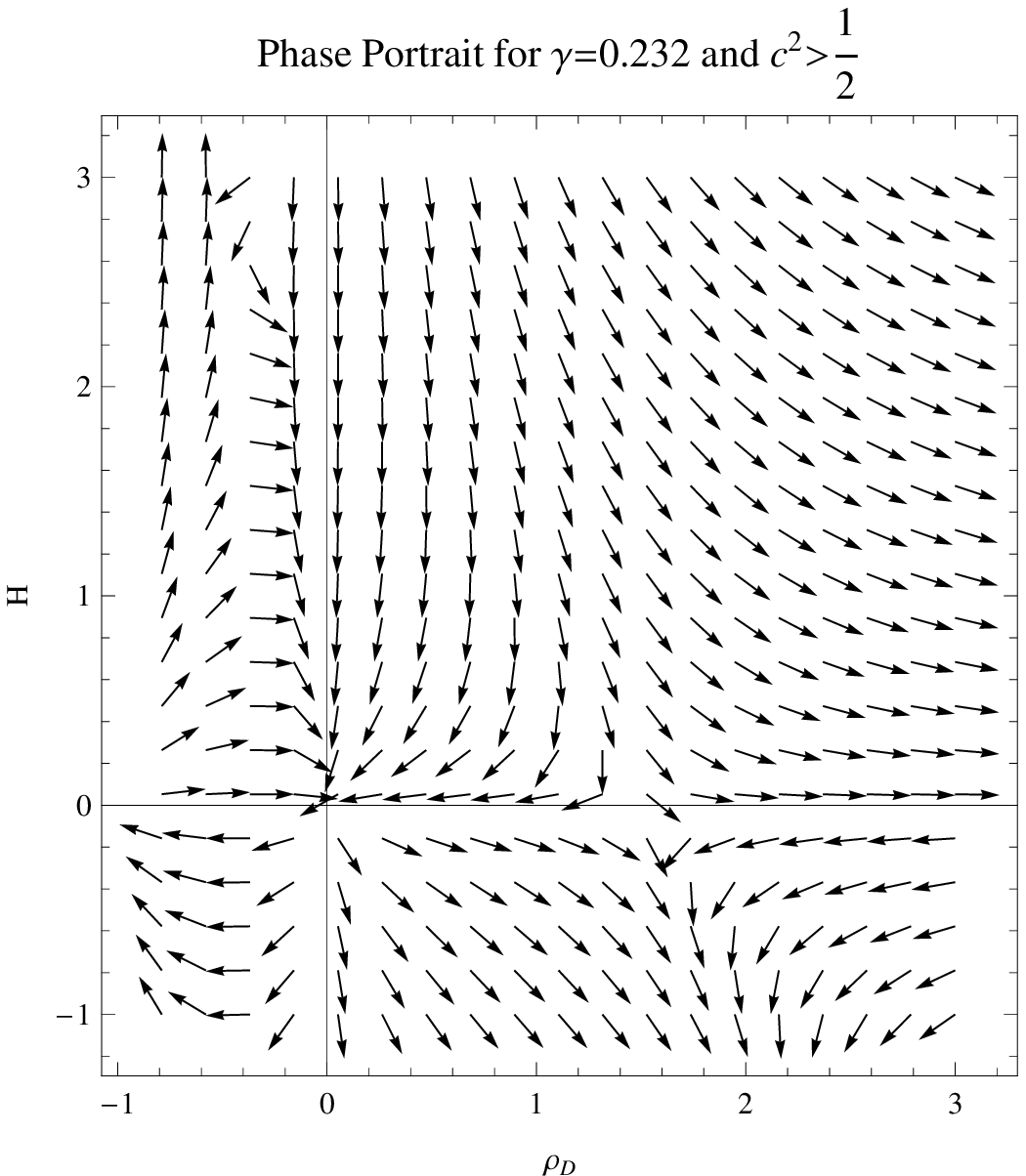}\\
\\
~~~~~~~~~~~~~~~~~~~~~~~~~~~~~~~~~~~~~~~~~~~~~~~~~~~~~~~~~~~~~~~~~Fig.3(a)
~~~~~~~~~~~~~~~~~~~~~~~~~~~~~~~~~~~~~~~~~~~~~~~~~~~~~~~~~~~~~~~~Fig.3(b)
\\\\

\includegraphics[height=2.6in, width=2.6in]{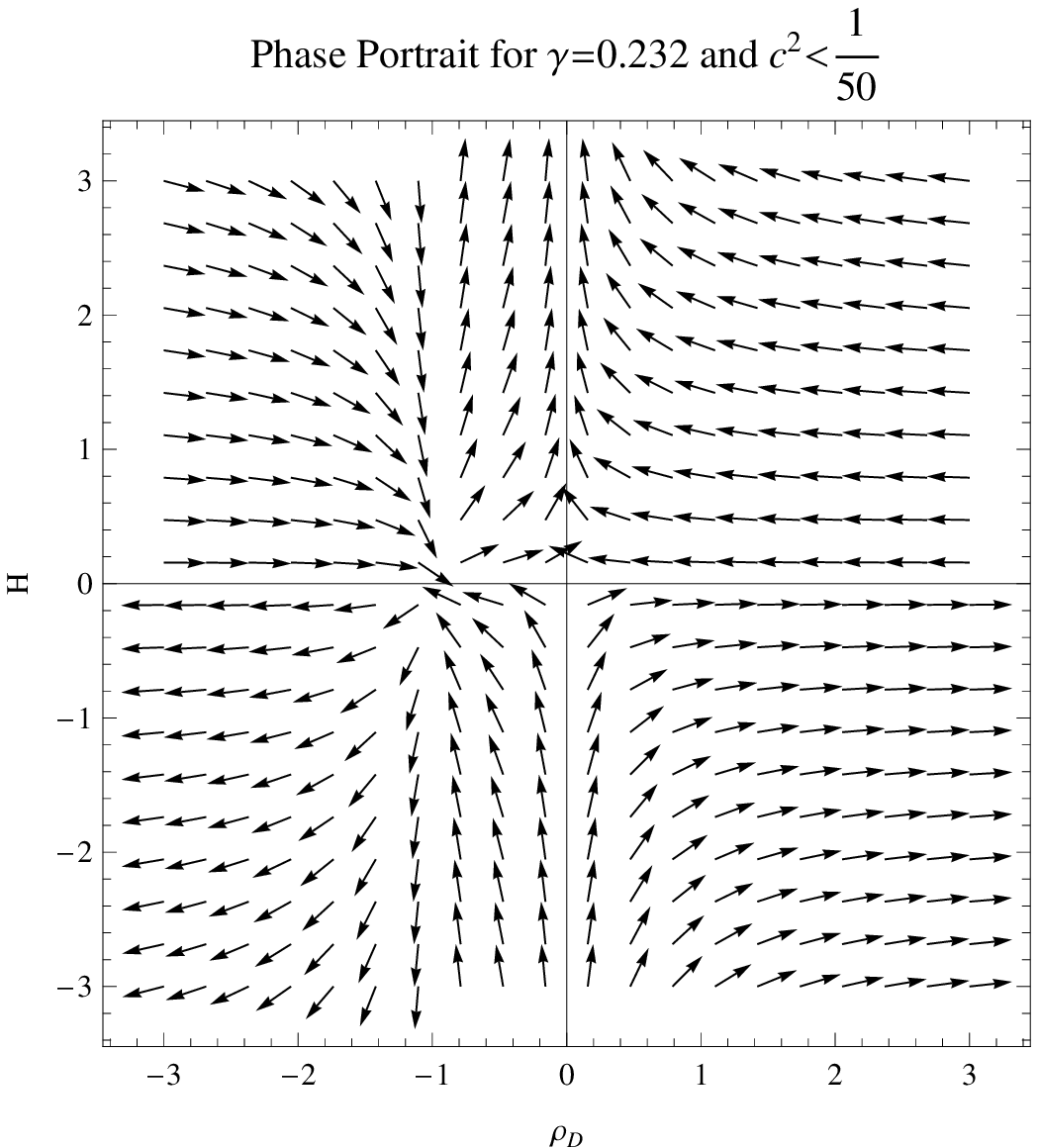}~~
\includegraphics[height=2.6in, width=2.6in]{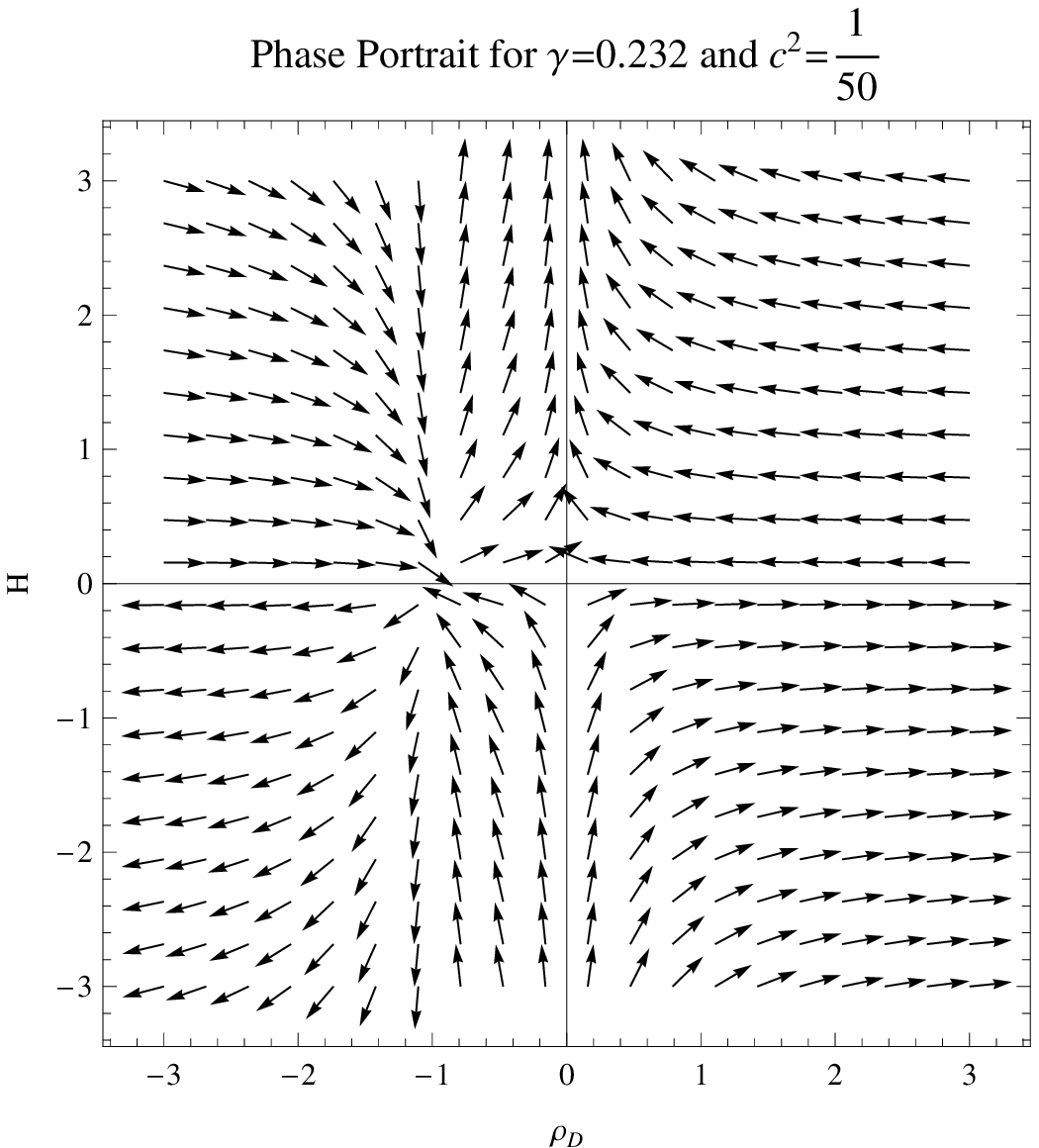}\\
\\
~~~~~~~~~~~~~~~~~~~~~~~~~~~~~~~~~~~~~~~~~~~~~~~~~~~~~~~~~~~~~~~~~Fig.3(c)
~~~~~~~~~~~~~~~~~~~~~~~~~~~~~~~~~~~~~~~~~~~~~~~~~~~~~~~~~~~~~~~~Fig.3(d)
\\\\

\includegraphics[height=2.6in, width=2.6in]{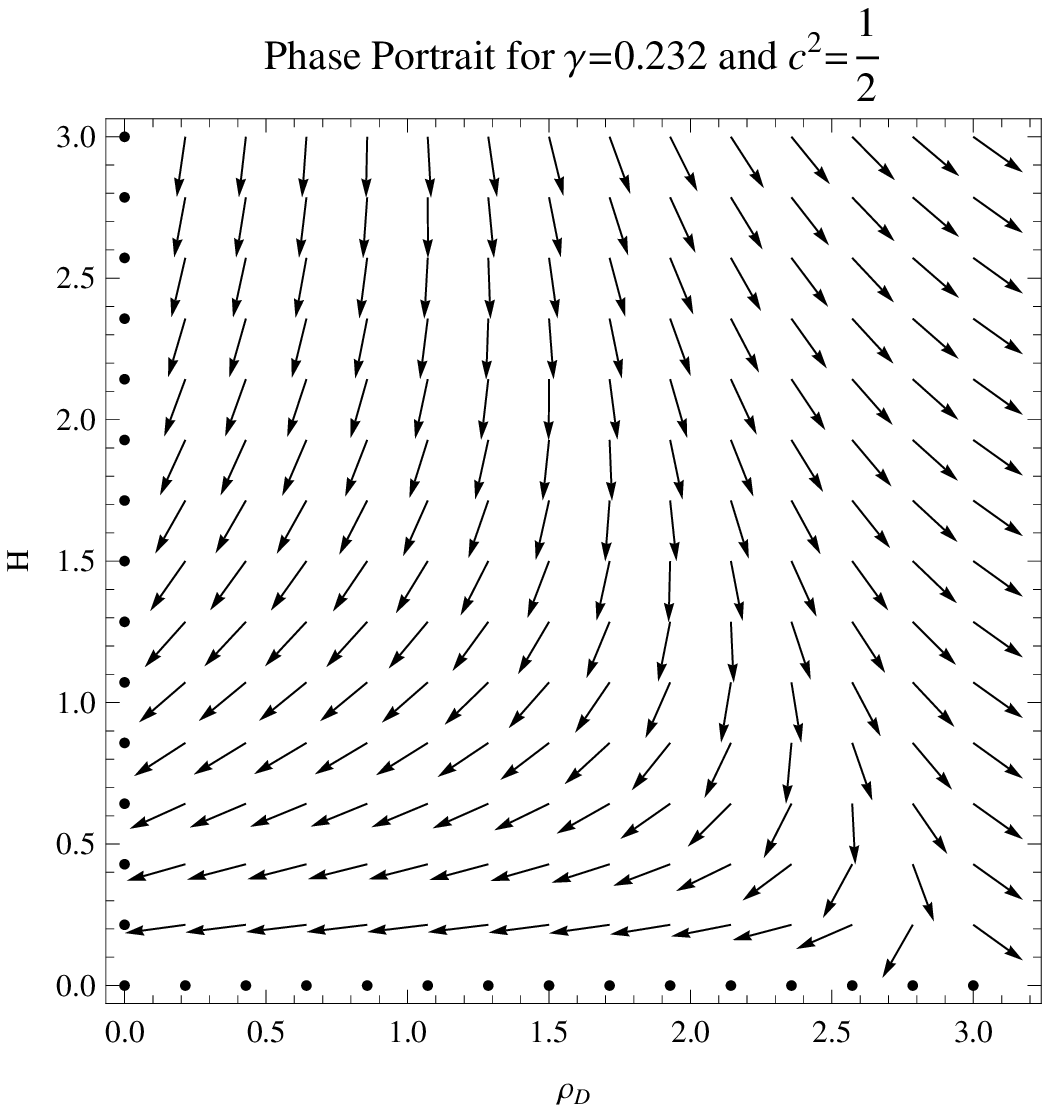}~~

~~~~~~~~~~~~~~~~~~~~~~~~~~~~~~~~~~~~~~~~~~~~~~~~~~~~~~~~~~~~~~~~~Fig.3(e)
\\\\

Fig.3(a)-3(e) represent the variation of $\Omega_D - H$. Here also
the negative coordinate of $H~and~\Omega_D$ is not physically
valid. But for completeness of the system we have drawn the whole
figure.These figures characterizes the nature of the critical
points given in table2.

\hspace{1cm} \vspace{2cm}

\end{figure}

\section{Discussion:}

The paper analyzes the HDE model interacting with DM(in the form
of dust). Here the IR cut off is chosen at the Ricci's length with
the justification that it corresponds to the size of the maximal
perturbation corresponding to formation of a black hole. The
interaction between the two fluids is either a linear combination
or in product form of the two energy densities of which the
product form is physically more variable one. In both the cases,
the evolution equations are transformed to an autonomous system
for which the nature of the critical points are presented in
tabular form and are graphically analyzed. Finally, the
coincidence problem is discussed by studying the evolution of the
energy ratio for the first case only.\\\\\\\\\

{\bf Acknowledgement :\\} RB and NM want to thank West Bengal
State Govt. and CSIR, India respectively for awarding JRF. Authors
are thankful to IUCAA, Pune as this work was done there during a
visit.

\end{document}